\documentclass[12pt]{spieman}  
\usepackage{amsmath,amsfonts,amssymb}
\usepackage{graphicx}
\usepackage{setspace}
\usepackage{tocloft}
\usepackage{comment}
\usepackage{color}
\usepackage{soul}

\title{Estimation of systematic error from bulk deformation of end test mass induced by photon calibrator for LIGO post-O5 gravitational wave projects}

\author[a,b,c,d\dag]{Daiki Tanabe}
\author[a]{Aloysius Niko}
\author[b,a]{Kun-Yao Chang}
\author[b,a,c,d]{Yuki Inoue}
\author[e]{Dripta Bhattacharjee}
\author[f]{Richard Savage}
\author[d]{Henry Tsz-King Wong}

\affil[a]{Center for High Energy and High Field Physics, National Central University, Taoyuan 32001, Taiwan}
\affil[b]{Physics Department, National Central University, Taoyuan 32001, Taiwan}
\affil[c]{Institute of Particle and Nuclear Studies (IPNS), High Energy Accelerator Research Organization (KEK), Tsukuba, Ibaraki 305-0801, Japan}
\affil[d]{Institute of Physics, Academia Sinica, Nangang, Taipei, 015011, Taiwan}
\affil[e]{Physics Department, Kenyon College, Gambier, OH 43022-9623, USA}
\affil[f]{LIGO Hanford Observatory, Richland, WA 99352, USA}

\cftpagenumbersoff{figure}
\cftpagenumbersoff{table} 

\begin{document} 
\maketitle

\begin{abstract}
We studied the bulk deformation of the end test masses (ETMs) of gravitational wave (GW) detectors caused by calibration devices named Photon Calibrator (Pcal). This bulk deformation is one of the source of calibration error since it induces non-linear displacement from the ideal pendulum motion of the ETM. Particularly, it is the dominant error source above 1500 Hz which is crucial to the studies of neutron star mergers. The displacement also depends on beam offsets of the main interferometer (IFO) beam and Pcal beams. We conceptually described frequency dependent effect of bulk deformation and evaluated it in several cases of the beam offset with finite-element analysis (FEA) software, for ETM models of Advanced LIGO (aLIGO), KAGRA, Advanced Virgo (AdVirgo), and LIGO A\#.  
\end{abstract}



{\noindent \footnotesize\textbf{\dag}Daiki Tanabe,  \linkable{tana2431.ts@gmail.com} }


\section{Introduction} \label{sect:intro_section}

Gravitational waves (GWs) have proven to be exceptional probes of fundamental physics, bridging the fields of physics and astronomy. Joint observations by Laser Interferometer Gravitational-wave Observatory (LIGO), the Advanced Virgo Observatory (Virgo), and the Kamioka Gravitational-wave Observatory (KAGRA) have detected more than a hundred gravitational wave events. These events originate primarily from binary mergers of black holes and neutron stars, providing a unique window into the most extreme conditions in the Universe.
The sensitivity and reach of these observations will be further enhanced by planned upgrades such as LIGO A\#, which aims to increase the sensitivity and refine the accuracy of the data obtained from these events. All of these international gravitational wave experiments use an advanced interferometer system designed to detect tiny distortions in spacetime caused by passing gravitational waves. 
Moreover, a global network of gravitational wave detectors, including the future LIGO India, enables reducing degeneracy of source parameters. It significantly enhances the accuracy of source localization and polarization. This expanded geographical spread not only improves detection rates but also refines our understanding of gravitational wave polarization and propagation characteristics.

Calibration uncertainties in gravitational wave (GW) detection play a crucial role in influencing the accuracy of measurements related to GW signals. These uncertainties primarily affect the determination of the absolute amplitude of the GW signals. One of the most critical aspects where calibration uncertainties have a substantial impact is in estimating the distance to the source of gravitational waves. Since the luminosity distance to the GW source directly influences the inferred physical properties of the source, such as mass and orbital characteristics, any errors in distance measurement can significantly skew these estimates.

Calibration uncertainties can also impair the accuracy of sky localization of the GW sources. Accurate determination of GW source coordinates is essential for follow-up observations with electromagnetic telescopes. This is particularly challenging when only a few detectors, often just three, are available to detect the GW. Sensitivity of each interferometer in the detector network has a directional dependence, which means that the precision of localization largely depends on the geometry and alignment of the detectors relative to the source. Calibration errors can, therefore, lead to less precise coordinate determination.
The effect of calibration uncertainties becomes even more pronounced in events with high signal-to-noise ratios (SNR). In such cases, although the angular resolution—-how well we can pinpoint the location of the source in the sky—-is generally better due to the clearer signal, it still remains vulnerable to distortions from calibration errors. This makes high-SNR events particularly sensitive to the fidelity of calibration.

In order to mitigate these issues, photon calibrators (Pcals) are employed in the recent global network of gravitational wave detectors. The Pcal uses a laser source whose beam power is calibrated to produce a measurable displacement of end-test mass (ETM), thereby allowing us to calibrate the detectors' response more accurately. This calibration method ensures that the interferometers measure the GW signals with greater accuracy, thus reducing the impact of calibration uncertainties on the overall data quality and scientific conclusions drawn from GW observations.

The Pcal has undergone significant evolution since its inception, where it first saw application at the GEO600 and the Glasgow 10~m interferometers. As a pioneering Pcal, this first-generation system employed photon pressure as a means to actuate the ETM by injecting a laser beam directly onto the mirror surfaces~\cite{Hild_2007}. This system faced deformation of the mirror caused by the photon force applied on its surface. We distinguish two types of deformation; (i) ``Local elastic deformation'' observed around the contact points of the laser, which typically aligned at the center of mass of a mirror and (ii) ``Bulk deformation'' of the whole solid body resulted from propagation of the stress of local deformation. These deformations affect to the calibration accuracy by inducing surface displacement. The local deformation was proven to change the displacement sensed by the interferometer by more than 10\% of the ideal free-mass motion at frequencies higher than 1~kHz~\cite{Pcal_local_deformation}. Other key systematic errors of Pcal include optical efficiency, the uncertainty of power standards, and rotation of ETM.

\begin{figure}
\begin{center}
\begin{tabular}{c}
\includegraphics[height=8.0cm]{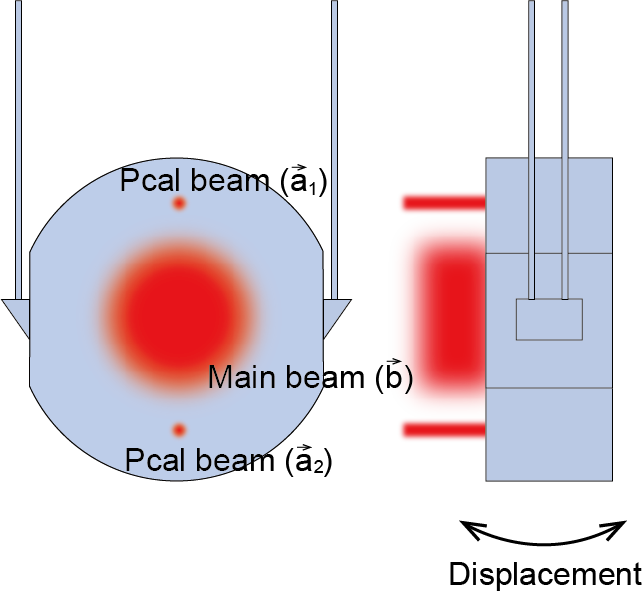}
\end{tabular}
\end{center}
\caption[]
{ \label{fig:Pcal}
Conceptual view of Pcal used in 4th observation run of LIGO. Pcal beam positions are expressed by position vectors $\vec{a}_1$ and $\vec{a}_2$, while the main interferometer beam is by $\vec{b}$.
}
\end{figure}

Addressing the local deformation issue, LIGO developed the second-generation Pcal system~\cite{karki2016advanced}. This system introduced the concept of two-points injection that located local deformation at two separate points outside the projection area of the main interferometer beam where the motion is sensed (Fig.~\ref{fig:Pcal}). This design mitigated the local deformation effect. Moreover, by optimizing the beam positions, it suppressed excitation of one of the fundamental bulk deformation modes, so called drumhead mode~\cite{Pcal_drumhead}. Additionally, LIGO integrated an optical-follower servo into this system. This servo acted to stabilize the laser power, thereby decreasing laser-induced noise and suppressing unwanted higher harmonic frequencies, enhancing the overall sensitivity and stability of the calibration system.

In the latest advancement, KAGRA has developed a third-generation Pcal system for the KAGRA observatory, in collaboration with LIGO~\cite{kagra_pcal}. This new iteration incorporated a 20~W continuous-wave laser, which raised the power output by an order of magnitude. Because response of pendulum is attenuated in proportion to the inverse square of frequency, increasing displacement by larger power was particularly essential for calibration at kHz region. An additional enhancement in the third-generation Pcal is the independent beam control system. This system was implemented to characterize the rotation effect caused by position offset and power imbalance of the beams. A crucial part of this system is the real-time monitoring of the beam position, which is vital for assessing any rotational effects and further bulk deformations that may occur. 

This paper focuses particularly on the systematic errors induced by bulk deformation at high frequency. Typical resonant peaks of bulk deformation are at a few kHz where a ringdown peak of a binary neutron star merger lies\cite{ligo_neutron_eos}. It can be a major source of systematic error in studies of equation of state of neutron stars. Quantitative estimations of the bulk deformation is essential for nuclear physics through GW experiments.

Previous studies of LIGO have reported methods to determine the optimal Pcal beam positions based on a nodal radius of the drumhead mode that is suppressed by the two-points injection scheme~\cite{sudarshan_phd}\cite{Nicola_pcal}. We extended these studies to the suspension structure containing the wires and penultimate mass above the test mass, for including translational motion and rotation into our evaluation. Calibration of LIGO in O4 used a rigid quadruple pendulum model to estimate response of the suspension and conservatively corrected the deformation effect~\cite{deformation_correction_o4}. We also applied the method for optimizing beam positions to the test-mass designs of other GW experiments. Furthermore, we evaluated the case that the beams have offset from the optimal positions, based on the measured beam positions in LIGO Hanford Observatory (LHO).

In the following part of this paper, Sec. 2 discusses a mathematical model for surface displacement. Section 3 details the simulation process using two types of simulation software, ANSYS and COMSOL, to confirm consistency of their results. Section 4 compares calculations for Advanced LIGO (aLIGO), Advanced Virgo (AdVirgo), KAGRA, and LIGO A\# experiments. Section 5 discusses difference among experiments caused by geometry and material of their ETMs.

\section{Model of displacement response}

The concept of degrees of freedom in an elastic body refers to the number of independent directions in which each point of the body can move. In a general three-dimensional space, a system consisting of $N$ particles has $3N$ spatial degrees of freedom. This means that each point can move along the x-axis, y-axis, and z-axis. In other words, any point can move in three directions, typically represented as three orthogonal directions. These degrees of freedom can be further categorized into translational, rotational, and internal modes. Translational and rotational degrees of freedom, as well as internal modes, represent different aspects of the motion and vibration of an object, each with independent characteristics that can influence each other around the resonant frequency. Each degree of freedom can be explained as follows:

\paragraph{Translational degrees of freedom}
Represent the ability of an object to move linearly in one or multiple dimensions within space. For example, movement in horizontal or vertical directions corresponds to this.
These degrees primarily represent changes in the position of the object.
\paragraph{Rotational degrees of freedom}
Represent the ability of an object to rotate about a point or axis. This includes changes in the object's orientation or angle.
These degrees represent how the object changes its orientation or rotates.
\paragraph{Internal modes}
Represent internal vibrations or inherent vibration modes of an object. This includes vibrations of elastic bodies, structural resonance modes, stress waves in materials, etc.
Internal modes depend on the structure of the object and material properties and can be excited by external forces or own motion of object.

\begin{figure}[tbp]
\begin{center}
\begin{tabular}{c}
\includegraphics[height=7.0cm]{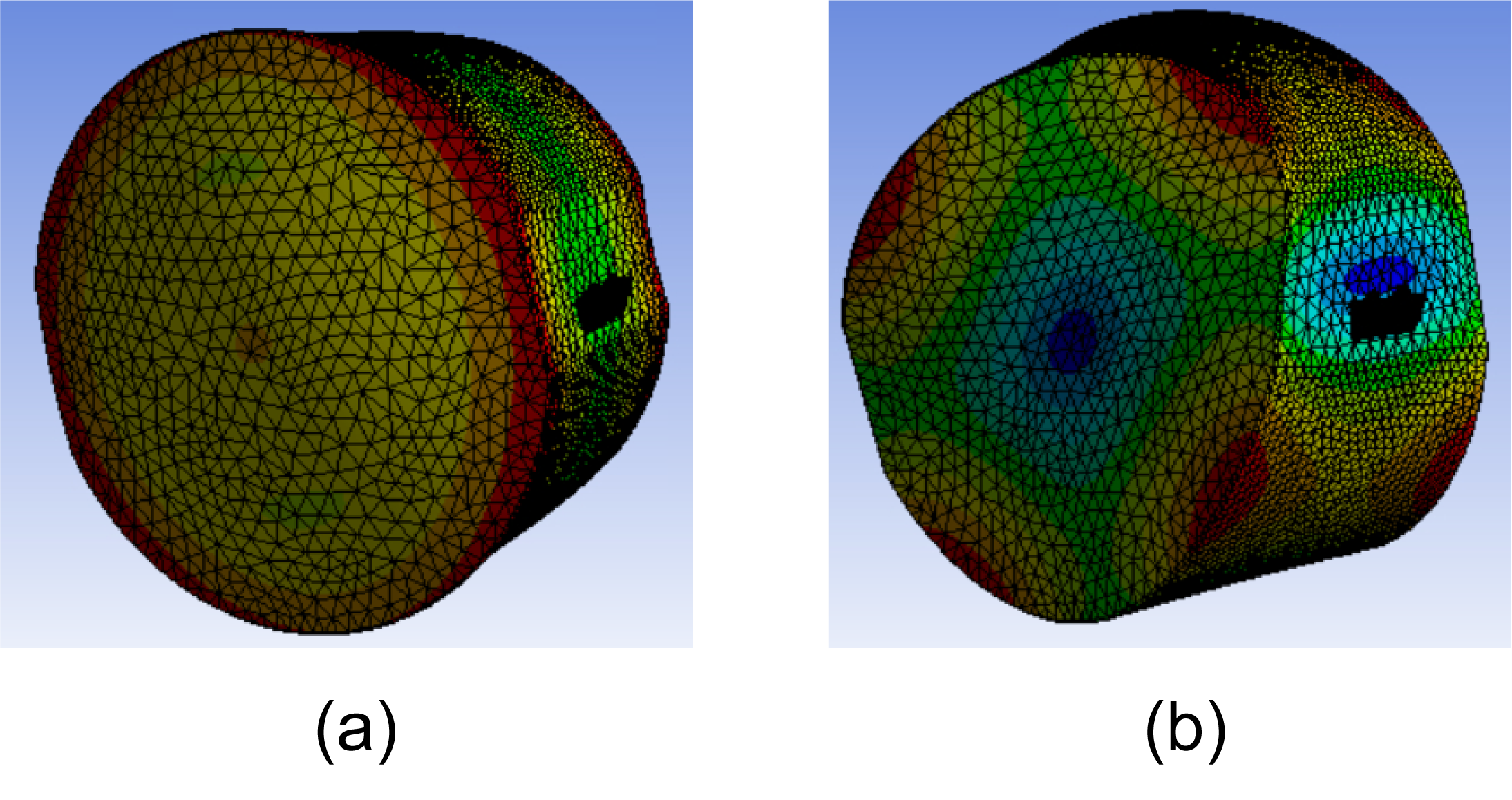}
\end{tabular}
\end{center}
\caption[]
{ \label{fig:twomodes}
Shapes of the two fundamental modes of a cylinder-like ETMs. The aLIGO's ETM is shown as an example. (a) Drumhead mode. (b) Butterfly mode. The color scale is arbitrary.}
\end{figure}

The movements of translation and rotation can induce excitation or change in internal vibrations. Internal modes can impact the overall response of the object and may provide feedback to translational and rotational movements. Particularly, when resonance occurs, it can significantly affect the motion and stability of the object.
Thus, translational, rotational, and internal modes each possess distinct dynamic characteristics as well as interactions each other. In this study, we disregard interactions as we focus on slight vibrations at points away from the resonance frequencies of internal modes. Internal modes discussed in this study consist of two types: drumhead mode and butterfly mode.
Drumhead mode and butterfly mode are types of vibration modes that occur when a circular membrane vibrates. In the cases of the cylinder-like objects, shapes of these modes slightly change by finite thickness and open boundary condition at the edges. Figure~\ref{fig:twomodes} shows the characteristic shapes of two modes in aLIGO's ETM. The followings are properties of drumhead mode and butterfly mode.

\paragraph{Drumhead mode}
Involves vibration in the direction of surface normal of the mirror in which the amplitude is point-symmetrical to the center. It is approximately characterized by index $n$ of the Bessel function $J_n$. In the lowest drumhead mode of a cylinder, the center and edges of the round surface vibrates in the opposite phase in the direction of surface normal. As a result, a nodal circle, where the displacement is always zero, appears on the surface. The edges also vibrate in the radius direction. Higher order drumhead modes tends to have multiple nodal circles, but they are suppressed because energy is consumed for expansion of the cylinder body. 

\paragraph{Butterfly mode}
Involves vibration in the direction of surface normal of the mirror in which the amplitude is symmetrical to the lines across the mirror surface. It is approximately characterized by index $m$ that indicates the order of zeros of the Bessel function $J_n$. In this mode, amplitude is minimum at the center and increases towards the outer edges. Nodes appear as lines across the center. Higher order butterfly modes have a larger number of node lines.

As we confirm in Sec.~\ref{sect:eigenresult}, the first drumhead mode of ETM has higher frequency than the first butterfly mode. Higher order modes are out of the observation band of the current GW detectors and rapidly decay by frequency. However, they slightly change the optimal position of the Pcal beams to suppress excitation of deformation. We determine the optimal beam positions in Sec.~\ref{sect:nodal}.

We consider the case where two points of laser light are injected on the mirror surface. Using the DC power $P_0^{(i)}$ and modulation $P^{(i)}_m(\omega)$, we define the incident power at the $i$-th point as $P^{(i)}=P_0^{(i)}+P^{(i)}_m(\omega)$. Then, the frequency-dependent force and torque exerted by the $i$-th laser on the mirror can be defined as $\vec{F}^{(i)}_m(\omega)=\frac{2P^{(i)}_m(\omega)}{c} \vec{n}^{(i)}$, $\vec{N}^{(i)}_m(\omega)=\vec{F}^{(i)}_m(\omega) \times \vec{a}^{(i)}$, respectively.

If we define the force-displacement transfer function and the force-rotation transfer function as $S_l=1/M\omega^2$ and $S_r=1/I \omega^2$ respectively, the displacement and rotation of each component can be expressed as:

\begin{eqnarray}
\vec{x}^{(i)}(\omega)=\vec{F}^{(i)}_m(\omega) S_l(\omega) \\
\vec{\theta}^{(i)}(\omega)=\vec{N}^{(i)}_m(\omega) S_r(\omega)
\end{eqnarray}

The displacement in the optical axis direction is practically measured by differential arm length (DARM) of the interferometer. When considering these values, three effect on the length must take into account: translational, rotational, and of elastic deformation. For this purpose, we define the normal vector $\vec{n}_l$ to the mirror surface. The translational effect of each component can then be written as:
\begin{eqnarray}
x_l(\omega) &=& \sum_i \vec{x}^{(i)}(\omega) \cdot \vec{n}_l \\
&=& \sum_i \frac{2P^{(i)}_m(\omega)}{c} S_l(\omega) \vec{n}_i \cdot \vec{n}_l \\
&=& \sum_i \frac{2P^{(i)}_m(\omega) \cos{\phi_i}}{c} S_l(\omega) 
\end{eqnarray}
 the rotational effect for each component, assuming the main beam position vector $\vec{b}=(b_x,b_y)$ is defined as:
\begin{eqnarray}
\label{eq:xr}
x_r(\omega)&=&\sum_i(\vec{\theta}^{(i)}(\omega) \times \vec{b}) \cdot \vec{n}_l \\
&=&S_r(\omega) \sum_i \left\{ \left( \vec{F}^{(i)}_m(\omega) \times \vec{a}^{(i)} \right) \times \vec{b}\right\} \cdot \vec{n}_l \\
&=&S_r(\omega) \sum_i \left\{(\vec{a}^{(i)} \cdot \vec{b})\vec{F}_m^{(i)}(\omega)-(\vec{a}^{(i)} \cdot \vec{F}_m^{(i)}(\omega))\vec{b} \right\} \cdot \vec{n}_l \\
&=& S_r(\omega)  \sum_i (\vec{a}^{(i)} \cdot \vec{b})\vec{F}_m^{(i)}(\omega) \cdot \vec{n}_l \\
&=& \sum_i (\vec{a}^{(i)} \cdot \vec{b}) \frac{2P^{(i)}_m(\omega)}{c} S_r(\omega) \vec{n}_i \cdot \vec{n}_l \\
&=& \sum_i (\vec{a}^{(i)} \cdot \vec{b}) \frac{2P^{(i)}_m(\omega) \cos{\phi_i}}{c} S_r(\omega) 
\end{eqnarray}
Finally, the effect of elastic deformation on the mirror surface coordinates $(\xi,\eta)$ with the Gaussian beam intensity $G(\xi,\eta;\vec{a})$ and deformation $D(\xi,\eta;\omega) \vec{n}_l$, is:

\begin{eqnarray}
\label{eq:x_e}
x_e(\omega)&=& \frac{\iint d\xi d\eta G(\xi-b_x,\eta-b_y) D(\xi,\eta,\vec{a};\omega) }{\iint d \xi d \eta G(\xi-b_x,\eta-b_y)}
\end{eqnarray}

We can define the transfer-like function by normalizing $x_e$ by power as
\begin{equation}
   S_e(\vec{a},\vec{b};\omega) =  \left( \sum_i \frac{c}{2P^{(i)}_m(\omega) \cos{\phi_i}}\right)  \cdot \frac{\iint d\xi d\eta G(\xi-b_x,\eta-b_y) D(\xi,\eta,\vec{a};\omega) }{\iint d \xi d \eta G(\xi-b_x,\eta-b_y)} 
\end{equation}

The total displacement is thus:
\begin{eqnarray}
x(\omega) &=& x_l(\omega) + x_r(\omega) + x_e(\omega) \\
&=&  \sum_i \frac{2P^{(i)}_m(\omega) \cos{\phi_i}}{c} (S_l(\omega)+S_r(\omega) \vec{a}^{(i)} \cdot \vec{b})+\left(\sum_i \frac{2P^{(i)}_m(\omega) \cos{\phi_i}}{c}\right)  S_e(\vec{a}^{(i)},\vec{b};\omega) )
\end{eqnarray}

Rotation and internal modes dominate different frequency range. Normalizing by the $1/M\omega^2$ factor of the translational motion, rotation is frequency-independent while internal modes rise toward the first butterfly mode. Consequentially, rotation dominates in O(10)-O(1000) Hz region and internal modes dominate above kHz region. We simulate these frequency dependencies in Sec.~\ref{sect:result}.

\section{Simulation configurations}
\label{sect:simulation}

\subsection{Two simulation software: COMSOL Multiphysics and ANSYS Mechanical}
We simulated bulk deformation assuming ETMs and Pcal configurations of four GW experiments; aLIGO, AdVirgo, KAGRA, and LIGO A\#. LIGO A\# is one of a future proposal of LIGO which plans to deploy a larger ETM~\cite{post_O5,ligo_asharp}. Bulk deformation was simulated by two finite-element analysis (FEA) software; COMSOL Multiphysics 5.4 (hereafter COMSOL) and ANSYS Mechanical r2019 (hereafter ANSYS). We mainly used COMSOL and crosschecked its results with ANSYS. Previous studies from LIGO and KAGRA reported consistency between these two software~\cite{sudarshan_phd}\cite{kagra_pcal}.

We adopted the same parameters for materials and Pcal configurations over all simulations as summarized in Tab.~\ref{tab:parameters}. Geometry of LIGO A\# were derived by scaling up the aLIGO ETM to be 100~kg with the same material. This scaling made it 1.36 times larger in each direction. Pcal beam radii were only used in ANSYS Mechanical. All Pcal injection angles were assumed to be $0^\circ$. The effect of injection angle is small enough because the results of frequency domain study is normalized by the amplitude of force. 
%
\begin{table}[tbp]
\centering
    \begin{tabular}{c|llll}
    \hline
    Parameter(unit) & aLIGO~\cite{sudarshan_phd} & AdVirgo~\cite{virgo_pcal} & KAGRA~\cite{kagra_pcal}~\cite{gwtw_2017_inoue} & LIGO A\#~\cite{post_O5,ligo_asharp}\\ 
    \hline \hline
    Diameter (mm) & 340 & 350 & 220 & 462.930 \\ 
    Thickness (mm) & 200 & 220 & 150 & 272.312 \\
    Weight (kg) & 39.618 & 42.37 & 22.994 & 100 \\
    Materials & Fused silica & Suprasil312 & Sapphire & Fused silica \\
    Density (kg/${\rm m^3}$) & 2203 & 2200 & 4000 & 2203 \\
    Poisson's ratio & 0.1631 & 0.17 & 0.3 & 0.1631 \\
    Young's modulus (GPa) & 72.6 & 70 & 400 & 72.6 \\
    Main beam radius (mm) & 62 & 58 & 35.3 & 62 \\
    Pcal beam radius (mm) & 2 & 2 & 3.491 & 2 \\
    Pcal power per beam (W) & 1 & 1 & 10 & 1 \\
    \hline
    \end{tabular}
\caption{Parameters used in simulations of bulk deformation.}
\label{tab:parameters}
\end{table}

\begin{figure}[tbp]
\begin{center}
\begin{tabular}{c}
\includegraphics[height=7.0cm]{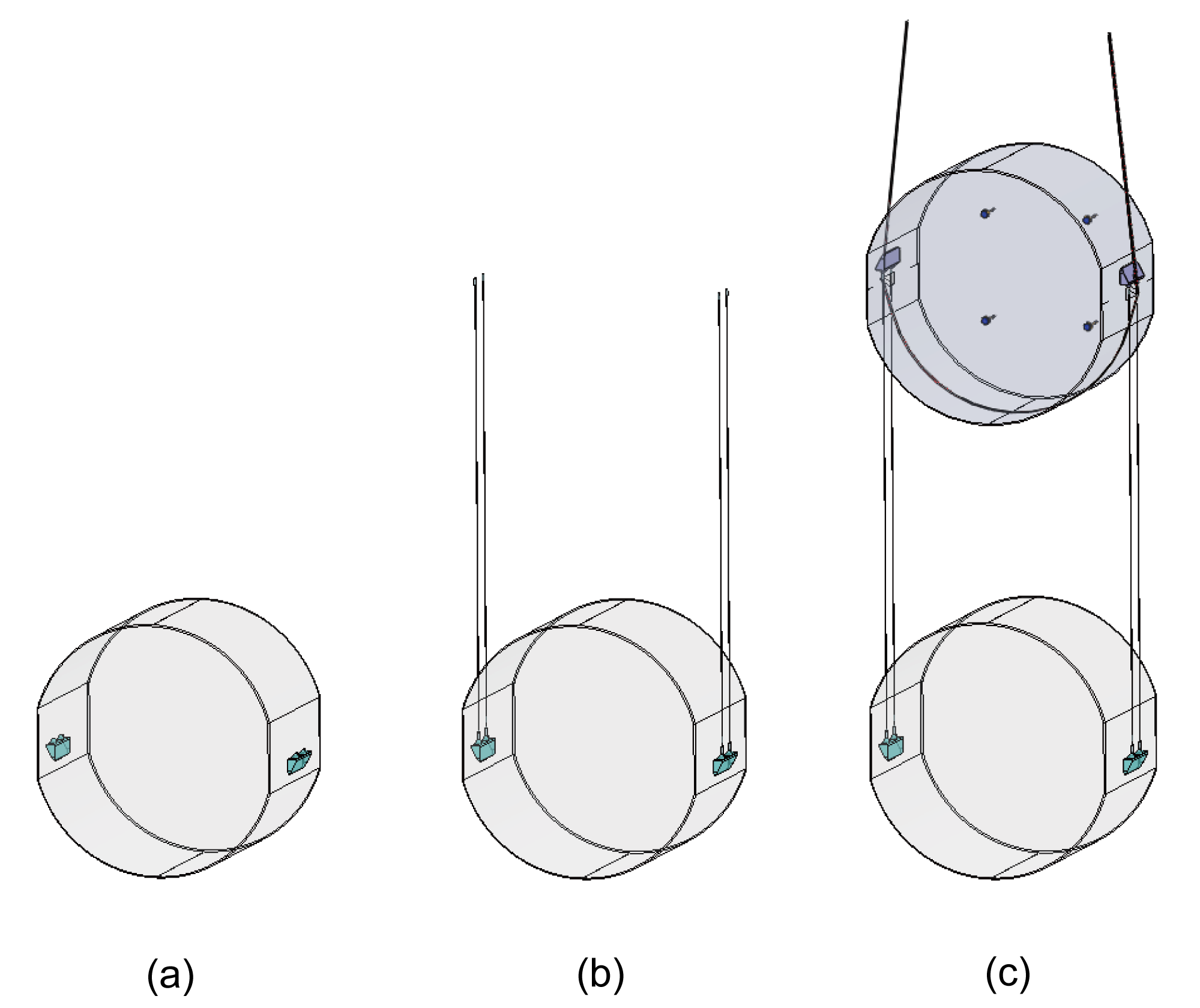}
\end{tabular}
\end{center}
\caption[]
{ \label{fig:suspensions}
Schematic view of (a) free-mass model (b) single suspension model (c) two-staged suspension model of aLIGO ETM.}
\end{figure}

We firstly used a ``free-mass" ETM models which did not include suspension structure so that were not constrained at anywhere. In addition, we simulated a single suspension model and a two-staged suspension model of aLIGO. As shown in Fig.~\ref{fig:suspensions}, The single suspension model includes four fused silica fibers suspending the ETM, and the two-staged suspension model includes the four fibers and a penultimate mass above. Although these suspension models enable us to simulate more realistic responses including rotation, they require relatively large computing resources in turn. We compared simulation results provided by the free-mass model and suspension models to check validity of using the free-mass model.

Our simulation flow consists of two processes; eigenfrequency study (modal analysis in ANSYS's term) and frequency domain study (harmonic response analysis in ANSYS's term). Eigenfrequency study in COMSOL performs modal analysis and finds eigenfrequencies and shapes of internal deformation of ETMs. It was used to predict the frequencies where the effect of butterfly mode deformation rises and to estimate nodal radius where portions of a ETM surface don't move so that is suitable to inject Pcal lasers. 

Frequency domain study applies external harmonic forces and simulates displacement of each point of ETM in a physical unit. Therefore, it is a direct estimation of bulk deformation. It is used for estimating the optimal Pcal beam positions where the total bulk deformation is minimized. Since higher order internal modes than butterfly and drumhead modes contribute to the total deformation, the optimal positions are not exactly the same as the nodal radius of drumhead mode. Furthermore, it allows us to evaluate excess of bulk deformation when the Pcal and interferometer beams have offset from their optimal positions. From a perspective of estimation of calibration error, it is convenient to evaluate bulk deformation by ratio to the rigid mass transfer function, i.e. $S_e(\vec{a},\vec{b};\omega)/S_l$. Here we call it displacement ratio. For a perfectly rigid material, displacement becomes to 1. Discrepancy of displacement ratio from 1 corresponds to a fractional calibration error relative to the rigid mass assumption. Here we define this error by $\Delta d\equiv S_e(\vec{a},\vec{b};\omega)/S_l -1$. 

\subsection{Meshing}
On COMSOL, we used an adaptive meshing spanning from 3.5~mm to 28~mm for masses of aLIGO, AdVirgo, KAGRA, and LIGO A\#. We specifically used 0.5~mm mesh for fibers in the suspension models of aLIGO. On ANSYS, we set mesh size on ETM surface to 4 mm.

\subsection{Eigenfrequency analysis}
We searched 24 modes from the lowest frequency. After computing modal analysis, we exported a list of displacement in the length direction at each mesh node on an ETM surface at a drumhead eigenfrequency for analysis to estimate a nodal radius. We extracted meshes in a vertical belt-like area of 1~cm width around the center line of an ETM. Then we detected the point where the absolute value of deformation is minimum in each region, and defined their vertical coordinates as nodal radius.

\subsection{Response in Frequency Domain}
Harmonic responses were studied at every 300 Hz in the range from 300 Hz to 10200 Hz for aLIGO, AdVirgo, LIGO A\# and every 500 Hz in the range from 500 Hz to 12000 Hz for KAGRA. This difference was for considering relatively higher eigenfrequencies of sapphire.

We applied Pcal-like pushing force on points defined on an ETM surface. On ANSYS, we defined a ``Pinball region" of the Pcal beam radius as injection points. This enables us to apply force in finite area. Positions of the injection points were determined based on nodal radius of drumhead mode. Amplitude of force was determined by $2P\cos\phi/c$. 

Since current AdVirgo injects one Pcal beam at the center of its ETM, we simulated both cases of one Pcal injection and possible two injections for AdVirgo. 

After computing frequency domain study, we evaluated displacement caused by bulk deformation. It was calculated by Based on Eq.~(\ref{eq:x_e}). By definition, it only considers amplitude of response and neglects phase. We integrated the surface displacement with a Gaussian profile of the main interferometer beam as a weight at each mesh, and then normalized it by the integrated main beam profile, net force of Pcal. The main beam profile was defined as an analytical function on COMSOL. The integration range was determined to be 30 cm square for aLIGO, AdVirgo, LIGO A\# while it was 20 cm square for KAGRA. The integrated main beam profiles in denominator were externally calculated in advance. On ANSYS, we exported the surface displacement at each mesh node as a list and calculated displacement ratio by python. The main beam profile and the integration range were defined in this post-simulation processing of ANSYS result.

\section{Simulation results with LVK-ETMs}
\label{sect:result}

\subsection{Simulated eigenfrequencies}
\label{sect:eigenresult}
We simulated eigenfrequencies of aLIGO ETM with COMSOL and ANSYS. In particular, we focused on the three lowest modes as main contributors to the displacement ratio in the GW observation frequency range. COMSOL and ANSYS provided consistent frequencies within 1 Hz. Eigenfrequency of the butterfly modes were at 5946 Hz and 6051 Hz, and the drumhead mode was at 8153 Hz. After consistency check of two FEA software, we simulated eigenfrequencies of ETMs for four GW experiments. The derived eigenfrequencies are summarized in Tab.~\ref{tab:all_eigenfrequency}. 
\begin{table}[tbp]
\centering
    \begin{tabular}{c|llll}
    \hline
    Mode & aLIGO & AdVirgo & KAGRA & LIGO A\# \\
    \hline \hline
    Butterfly (upright) (Hz) & 5946.0 / 5946.2 & 5616.5 & 15913.7 / 15913 & 4282.1 \\
    Butterfly (oblique) (Hz) & 6051.0 / 6051.0 & 5620.6 & 15978.6 / 15978 & 4357.1 \\
    Drumhead (Hz) & 8152.9 / 8152.9 & 7684.8 & 23658.7 / 23658 & 5901.1 \\
    \hline
    \end{tabular}
\caption{Eigenfrequencies of butterfly and drumhead modes of four ETMs. Values after slashes in aLIGO's and KAGRA's results show simulations by ANSYS. Other results are by COMSOL.}
\label{tab:all_eigenfrequency}
\end{table}

\subsection{Nodal radius of drumhead mode}
\label{sect:nodal}
\begin{figure}
\begin{center}
\begin{tabular}{c}
\includegraphics[height=6.0cm]{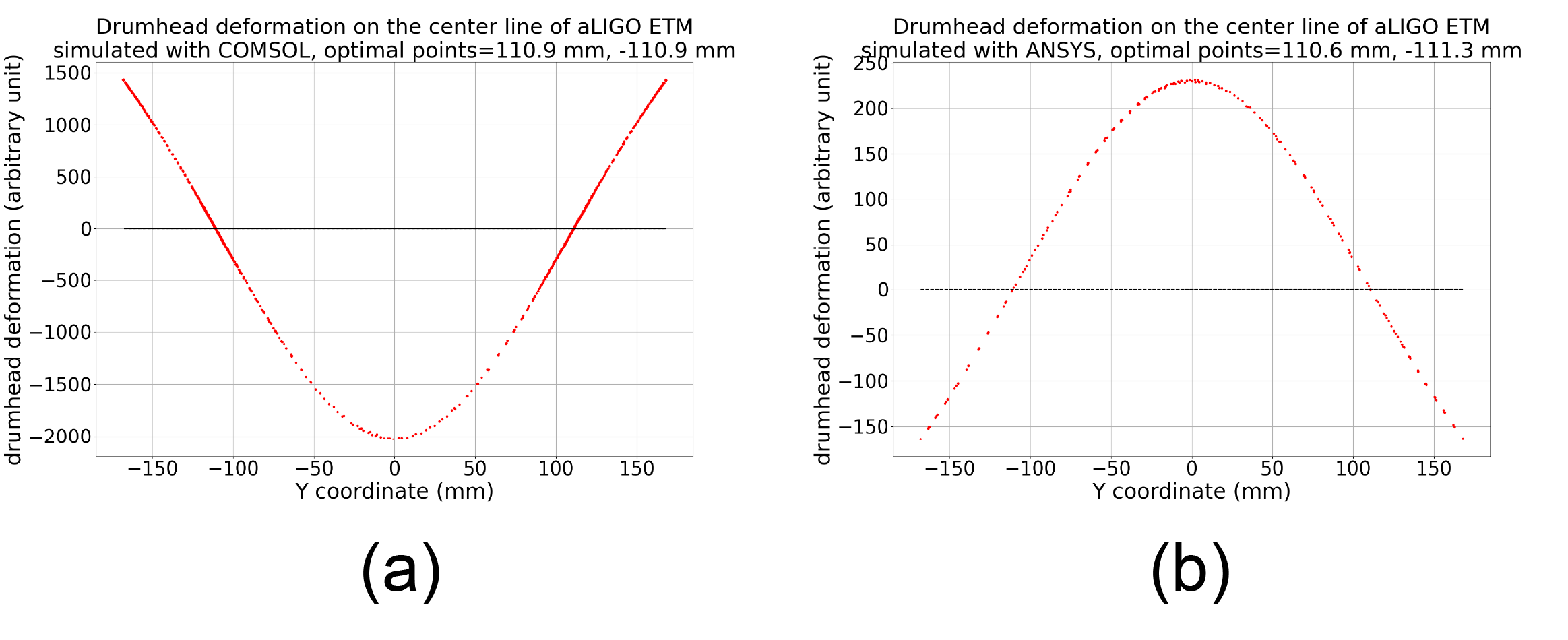}
\end{tabular}
\end{center}
\caption[]
{ \label{fig:aligo_drumhead_node} 
Drumhead mode shape on the vertical center line of aLIGO ETM. Polarity of convex depends on the shape at the final step of each simulation. (a) COMSOL simulation. (b) ANSYS simulation.}
\end{figure}

We processed the simulated displacement data of drumhead modes by python to know the nodal radius of the drumhead mode. Figure~\ref{fig:aligo_drumhead_node} shows nodal radii at $x=0$ mm on aLIGO ETM surface simulated by COMSOL and COMSOL. With COMSOL, the nodal radius was $\pm$110.9 mm while it was -111.3 mm and 110.4 mm with ANSYS. Differences were within 1.1 mm, which were small enough to expect that the optimal Pcal injection points are near this radius. Phase of drumhead oscillation shown in two software was opposite, but it does not affect to the eigenfrequency and the nodal radius.

We derived nodal radii of the other experiments by the same analysis as summarized in Tab.~\ref{tab:all_nodal}. Based on simulation by COMSOL, they were (-112.5 mm, 115.5 mm) for AdVirgo, (-66.4 mm, 62.8 mm) for KAGRA, (-151.6 mm, 152.0 mm) for LIGO A\#. Difference between COMSOL and ANSYS were all less than 1 mm.
\begin{table}[tbp]
\centering
    \begin{tabular}{c|llll}
    \hline
     Nodal radius & aLIGO & AdVirgo & KAGRA & LIGO A\# \\
    \hline \hline
     by COMSOL (mm, mm) & (-110.9, 110.9) & (-112.5, 115.5) & (-66.4, 62.8) & (-151.6, 152.0) \\
     by ANSYS (mm, mm) & (-111.3, 110.4) & - & (-65.3, 62.5) & - \\
    \hline
    \end{tabular}
\caption{Nodal radii on the vertical center lines of four ETMs.}
\label{tab:all_nodal}
\end{table}

\subsection{Harmonic response with ideally aligned beams}
\label{sect:harmonic}

\subsubsection{Displacement ratio of aLIGO ETM}
We searched the optimal Pcal beam positions where the displacement ratio is the closest to 1 by symmetrically moving two injection points along vertical axis of an ETM within 9 mm. This movement range follows a previous study~\cite{sudarshan_phd}. Here we particularly focus on 2.7-4.2 kHz range in which the main peak of ringdown in binary neutron star merger is expected to be~\cite{ligo_neutron_eos}. This region is shaded in all figures of $\Delta d$.

\begin{figure}
\begin{center}
\begin{tabular}{c}
\includegraphics[height=8.0cm]{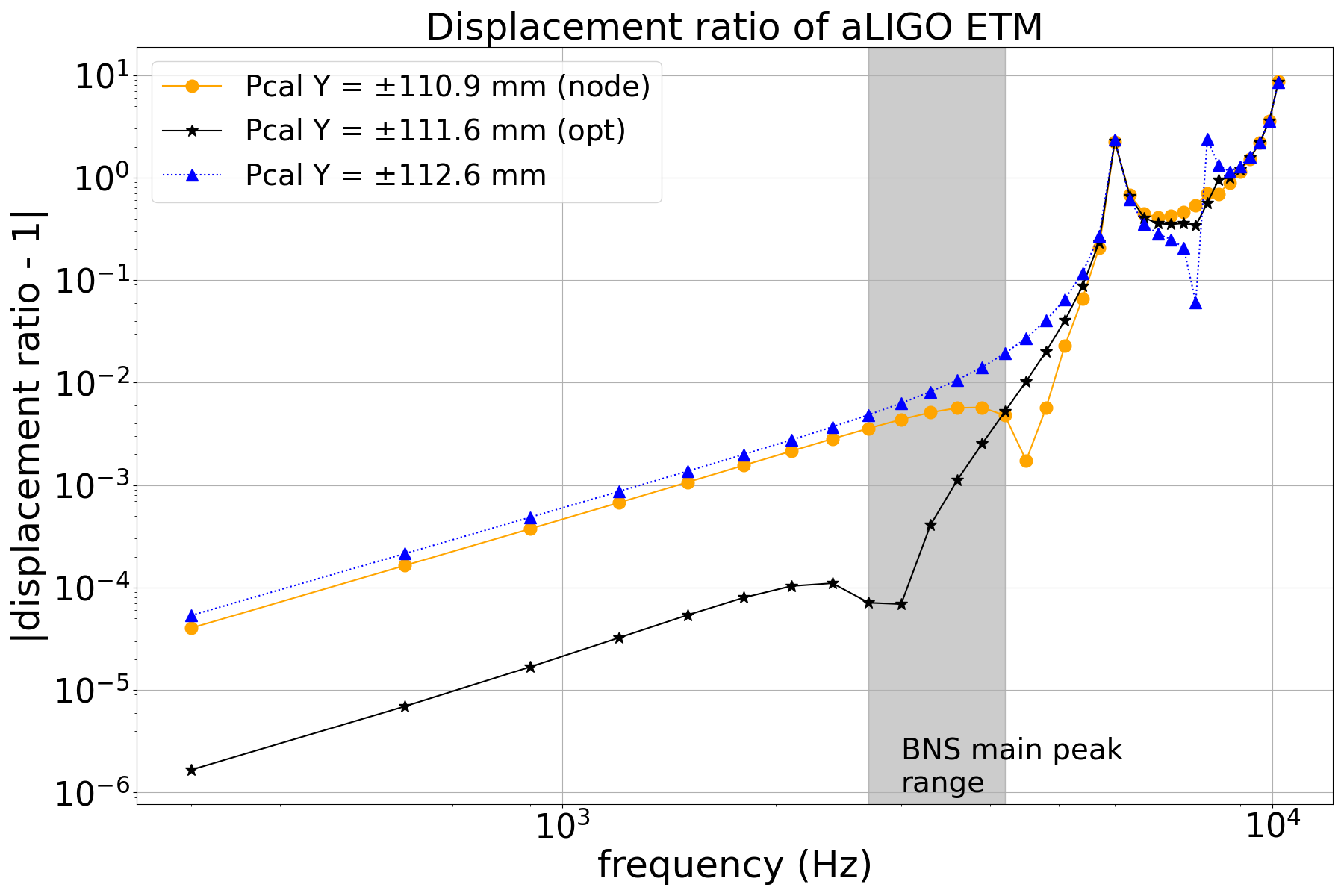}
\end{tabular}
\end{center}
\caption[]
{ \label{fig:aligo_loglog_shaded}
Discrepancy from the rigid mass motion of aLIGO ETM with several Pcal beam positions simulated by COMSOL.
({\it Solid Orange Circle}) Pcal beams are at the drumhead nodal radius,
({\it Solid Black Star}) Optimal position,
({\it Dashed Blue Triangle}) 1 mm outer from the optimal position.
The frequency range from 2.7 kHz to 4.2 kHz is shaded as the region of the main ringdown peak of binary neutron star merger.
}
\end{figure}

Figure~\ref{fig:aligo_loglog_shaded} shows comparison of displacement ratio of aLIGO ETM by Pcal beam positions. The lines rose by approaching to the eigenfrequencies of butterfly mode. Displacement ratio in the frequency range of interest becomes closest to 1 when the Pcal beams are distributed at $\pm$111.6 mm, 0.7 mm away from the estimation by nodal radius of drumhead mode. The error $\Delta d$ was -7.1e-5 at 2.7 kHz, and 5.2e-3 at 4.2 kHz. %

The fine-tuned optimal Pcal beam positions, discrepancy of displacement ratio from 1 at specific frequencies, and $\Delta d$ derived by COMSOL and ANSYS are summarized in Tab.~\ref{tab:disp_ideal}.
\begin{table}[tbp]
\centering
    \begin{tabular}{c|lllll}
    \hline
      & aLIGO & AdVirgo & AdVirgo & KAGRA & LIGO A\# \\
      &  & (one beam) & (two beams) &  &  \\
    \hline \hline
    Optimal Pcal beam position ($\pm$mm) & 111.6 & 0 & 116 & 75 & 147 \\
    $\Delta d$ at 2700 Hz & -7.1e-5 & -1.9e-0 & 9.7e-4 & -2.4e-4 & -2.4e-3 \\   
    $\Delta d$ at 4200 Hz & 5.2e-3 & -4.8e-0 & -4.6e-3 & -8.4e-4 & 2.1e-1 \\
    \hline
    \end{tabular}
\caption{Optimal Pcal beam positions tuned by frequency domain study and $\Delta d$ at the optimal positions simulated by COMSOL.}
\label{tab:disp_ideal}
\end{table}

\subsubsection{Comparison of COMSOL and ANSYS}
\begin{figure}
\begin{center}
\begin{tabular}{c}
\includegraphics[height=8.0cm]{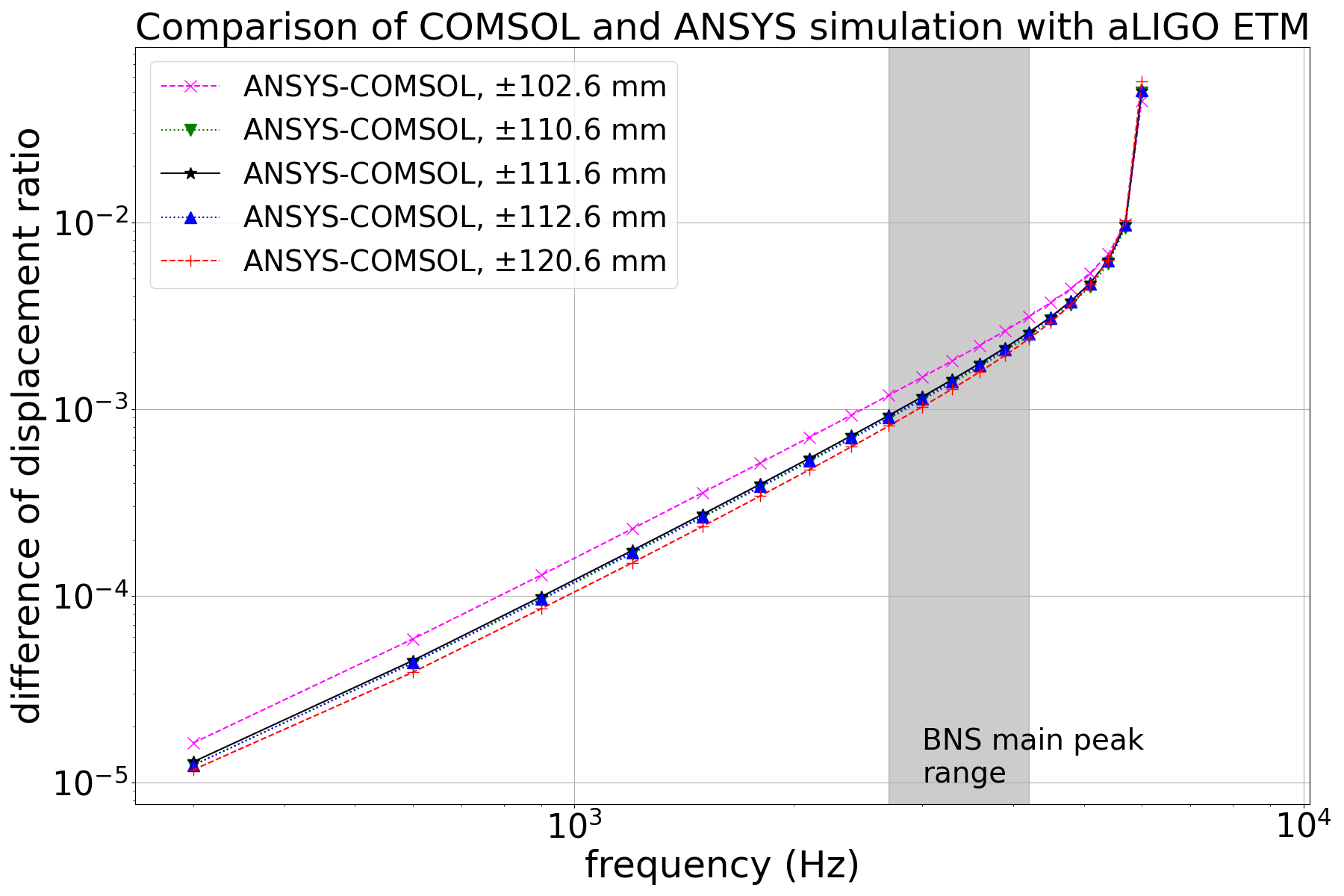}
\end{tabular}
\end{center}
\caption[]
{ \label{fig:aligo_ansyscomsol_loglog}
Difference between displacement ratio of aLIGO ETM with ideally aligned main beam simulated by COMSOL and ANSYS.
({\it Dashed Magenta Cross}) 9 mm inner from the optimal position,
({\it Dotted Green Inverted Triangle}) 1 mm inner from the optimal position,
({\it Solid Black Star}) Optimal position,
({\it Dotted Blue Triangle}) 1 mm outer from the optimal position,
({\it Dashed Red Plus}) 9 mm outer from the optimal position.
The frequency range from 2.7 kHz to 4.2 kHz is shaded as the region of the main ringdown peak of binary neutron star merger.
}
\end{figure}

Figure~\ref{fig:aligo_ansyscomsol_loglog} shows difference of displacement ratios for aLIGO ETM simulated by COMSOL and ANSYS. Ratio of displacement ratios from two software was 9.2e-4 at 2.7 kHz and 2.6e-3 at 4.2 kHz. Although the error rises near the resonant mode, it is still less than 1\%.

\subsubsection{Comparison of ETM and the perfect cylinder}
\label{sect:cylsus}
\begin{figure}
\begin{center}
\begin{tabular}{c}
\includegraphics[height=8.0cm]{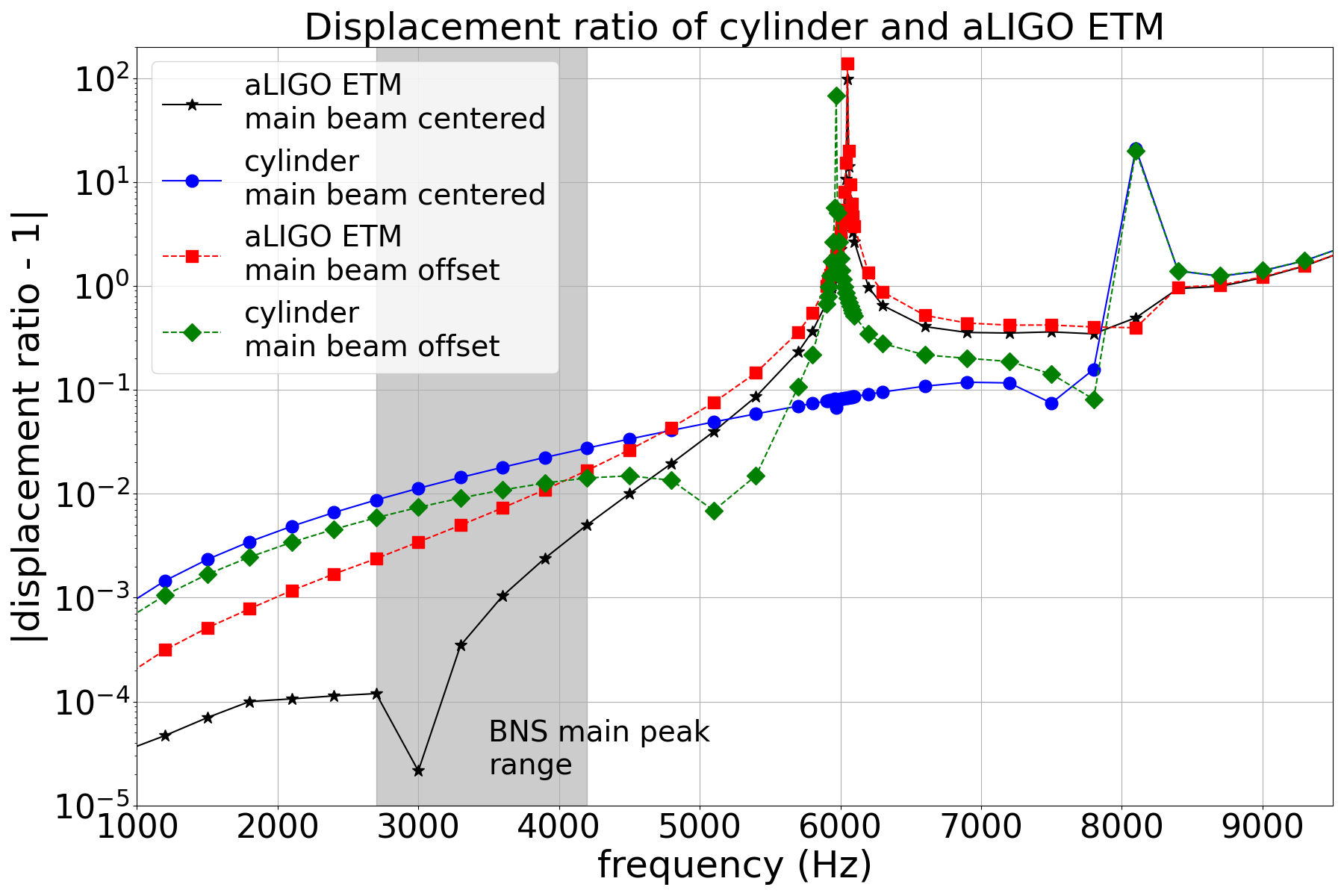}
\end{tabular}
\end{center}
\caption[]
{ \label{fig:aligo_cylsus}
Displacement ratio of the aLIGO ETM and perfect cylinder simulated by COMSOL.
({\it Solid Black Star}) aLIGO ETM with the centered main beam,
({\it Solid Blue Circle}) the perfect cylinder with the centered main beam,
({\it Dashed Red Square}) aLIGO ETM with the main beam offset,
({\it Dashed Green Diamond}) the perfect cylinder with the main beam offset.
The frequency range from 2.7 kHz to 4.2 kHz is shaded as the region of the main ringdown peak of binary neutron star merger.
}
\end{figure}

Next we compared displacement ratio of the actual aLIGO ETM and the perfect cylinder with the same diameter and thickness. Figure~\ref{fig:aligo_cylsus} shows the comparison of aLIGO ETM and the cylinder with- and without main beam offset. We assumed the offset of the main beam in the X-end of LHO which has been measured to be 16.2~mm to the right and 14.3~mm below the center of the ETM seen from the front~\cite{LHO_mainbeam}. The Pcal beams were assumed to be at the optimized positions for the aLIGO ETM and at the drumhead nodal radius for the cylinder. The difference of two shapes significantly appeared at the frequency near the deformation modes in kHz region. As approaching to the resonant frequency, the displacement ratio seen by the misaligned beam once dropped before the resonance. It was possibly due to an interaction between the two butterfly modes decomposed by the asymmetry of the mesh and ETM shape. The butterfly mode was not seen on the when the main beam is at the center of the perfect cylinder because the mode shapes are perfectly symmetric and their integration around the center is zero.

\subsubsection{Consistency between free-mass model and suspension models}
\label{sect:suspension}
\begin{figure}[tbp]
\begin{center}
\begin{tabular}{c}
\includegraphics[height=8.0cm]{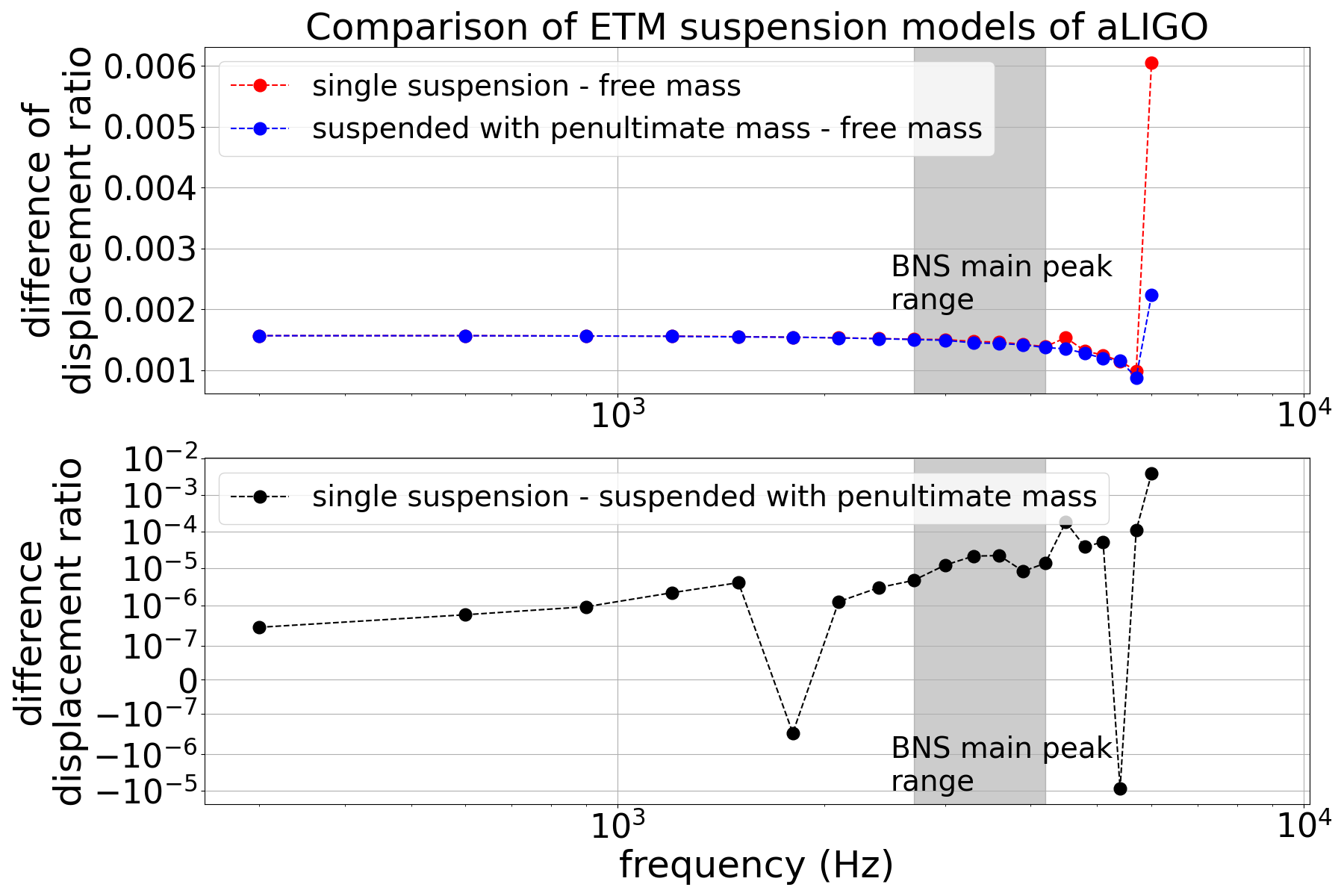}
\end{tabular}
\end{center}
\caption[]
{ \label{fig:freesusp}
Difference among displacement ratios calculated with the free-mass model, the single suspension model, and the two-staged suspension model of aLIGO ETM. ANSYS was used for simulation. 
({\it Red}) Difference between the single suspension model and the free-mass model,
({\it Blue}) Difference between the two-staged suspension model and the free-mass model,
({\it Green}) Difference between the single-suspension model and the two-staged suspension model.
The frequency range from 2.7 kHz to 4.2 kHz is shaded as the region of the main ringdown peak of binary neutron star merger.
}
\end{figure}

Finally we confirmed consistency between aLIGO's free-mass model, single suspension model, and two-staged suspension model. Figure~\ref{fig:freesusp} shows differences of displacement ratio when the Pcal beams are optimally aligned or vertically $\pm$9 mm away. We used ANSYS for these simulation since COMSOL failed to simulate the two-staged suspension model due to memory leak. Differences between the free-mass model and suspension models were 1.5e-3 at 2.7 Hz and 1.4e-3 at 4.2 kHz when the beams are at the optimal positions. The single suspension model and the two-staged suspension model differed for 4.7e-6 at 2.7 kHz and 1.4e-5 at 4.2 kHz.

\subsection{KAGRA, Virgo, LIGO A\#}
\begin{figure}
\begin{center}
\begin{tabular}{c}
\includegraphics[height=8.0cm]{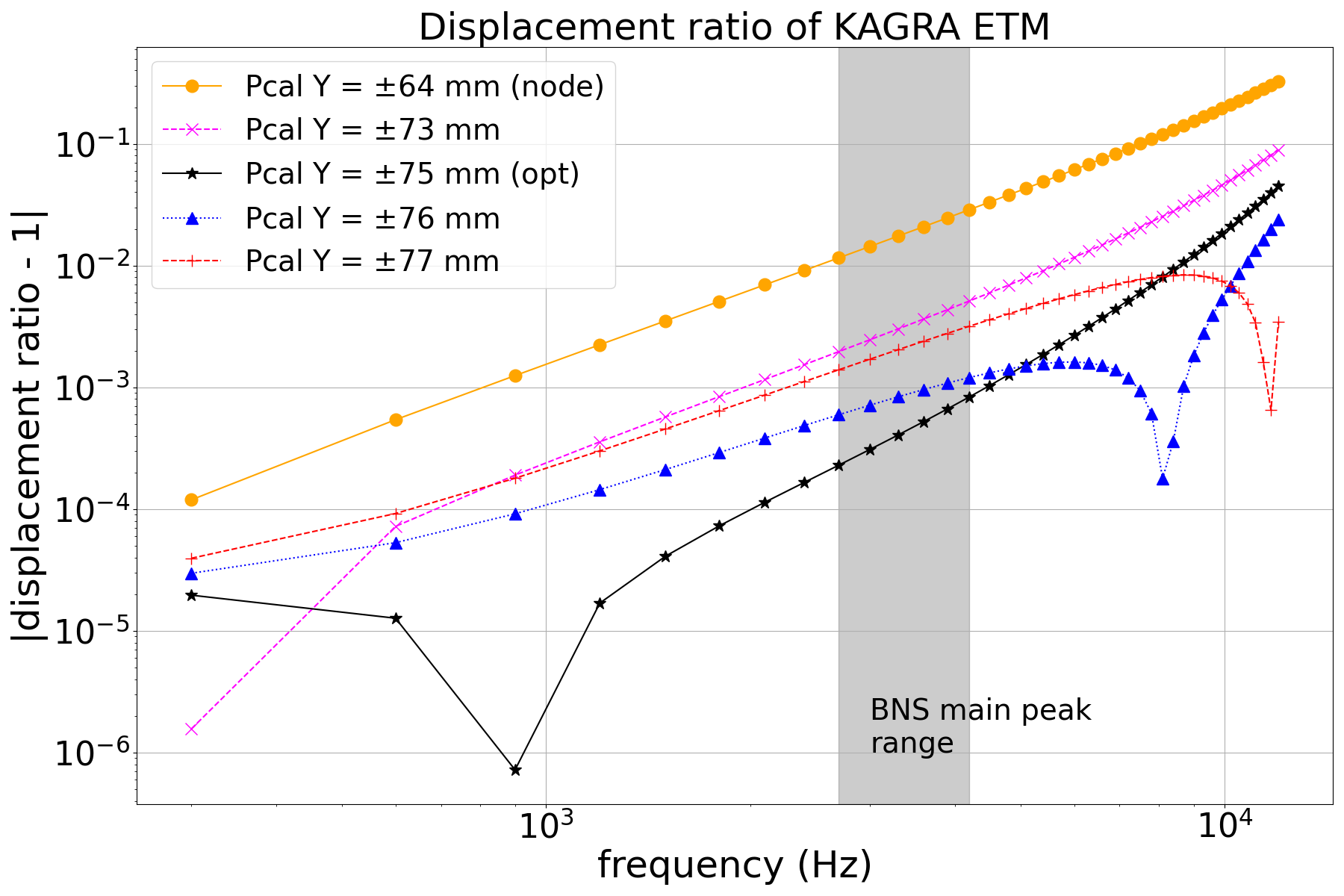}
\end{tabular}
\end{center}
\caption[]
{ \label{fig:kagra_loglog_shaded}
Discrepancy from the rigid mass motion of KAGRA ETM by various Pcal beam positions simulated by COMSOL.
({\it Solid Orange Circle}) Pcal beams are at the drumhead nodal radius,
({\it Dashed Magenta Cross}) 3~mm inner from the optimal position,
({\it Solid Black Star}) Optimal position,
({\it Dotted Blue Triangle}) 1~mm outer from the optimal position,
({\it Dashed Red Plus}) 2~mm outer from the optimal position.
The frequency range from 2.7~kHz to 4.2~kHz is shaded as the region of the main ringdown peak of binary neutron star merger.
}
\end{figure}

\begin{figure}[tbp]
\begin{center}
\begin{tabular}{c}
\includegraphics[height=8.0cm]{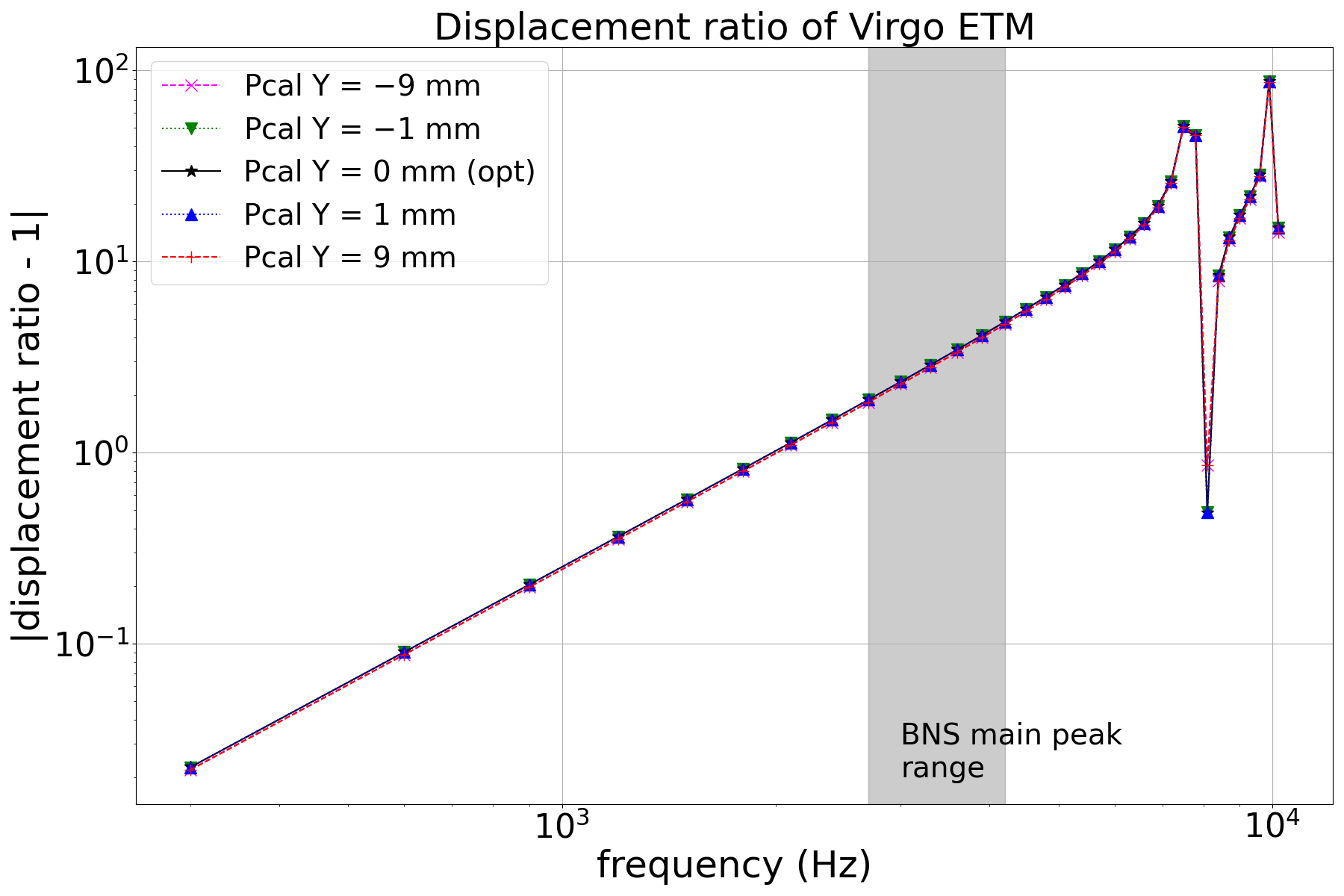}
\end{tabular}
\end{center}
\caption[]
{ \label{fig:virgo_center_loglog}
Discrepancy from the rigid mass motion of AdVirgo ETM in the single Pcal beam case simulated by COMSOL.
({\it Dashed Magenta Cross}) 9 mm lower from the optimal position,
({\it Dotted Green Inverted Triangle}) 1 mm lower from the optimal position,
({\it Solid Black Star}) Pcal beam is at the center,
({\it Dotted Blue Triangle}) 1 mm upper from the center,
({\it Dashed Red Plus}) 9 mm upper from the optimal position.
The frequency range from 2.7 kHz to 4.2 kHz is shaded as the region of the main ringdown peak of binary neutron star merger.
}
\end{figure}

\begin{figure}[tbp]
\begin{center}
\begin{tabular}{c}
\includegraphics[height=8.0cm]{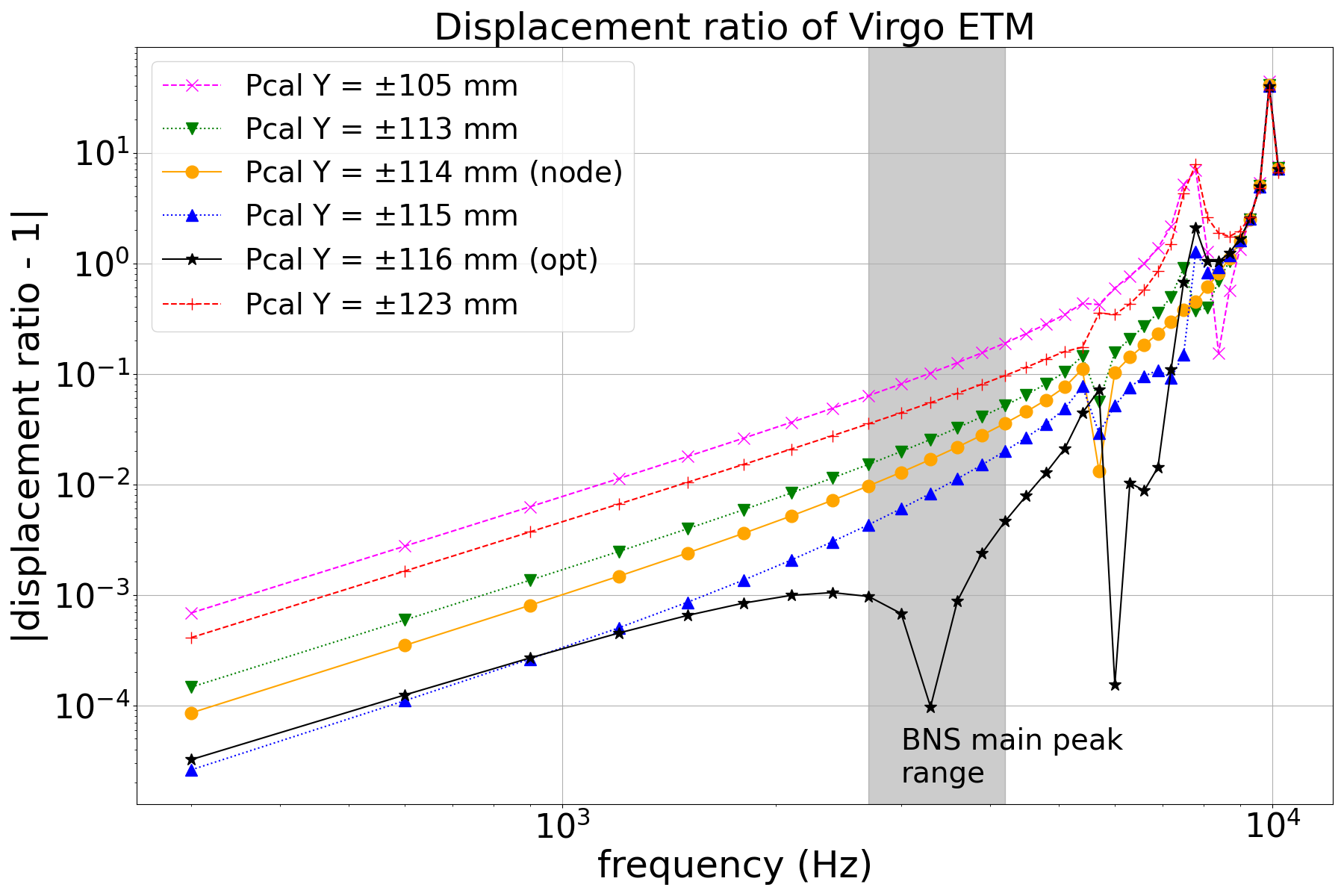}
\end{tabular}
\end{center}
\caption[] 
{ \label{fig:virgo_twobeam_loglog}
Discrepancy from the rigid mass motion of AdVirgo ETM in the two Pcal beams case simulated by COMSOL.
({\it Dashed Magenta Cross}) 9 mm inner from the drumhead nodal radius,
({\it Dotted Green Inverted Triangle}) 1 mm inner from the drumhead nodal radius, 
({\it Solid Orange Circle}) Pcal beams are at the drumhead nodal radius,
({\it Dotted Blue Triangle}) 1 mm outer from the drumhead nodal radius,
({\it Solid Black Circle}) Optimal position,
({\it Dashed Red PLus}) 9 mm outer from the drumhead nodal radius.
The frequency range from 2.7 kHz to 4.2 kHz is shaded as the region of the main ringdown peak of binary neutron star merger.
}
\end{figure}

\begin{figure}[tbp]
\begin{center}
\begin{tabular}{c}
\includegraphics[height=8.0cm]{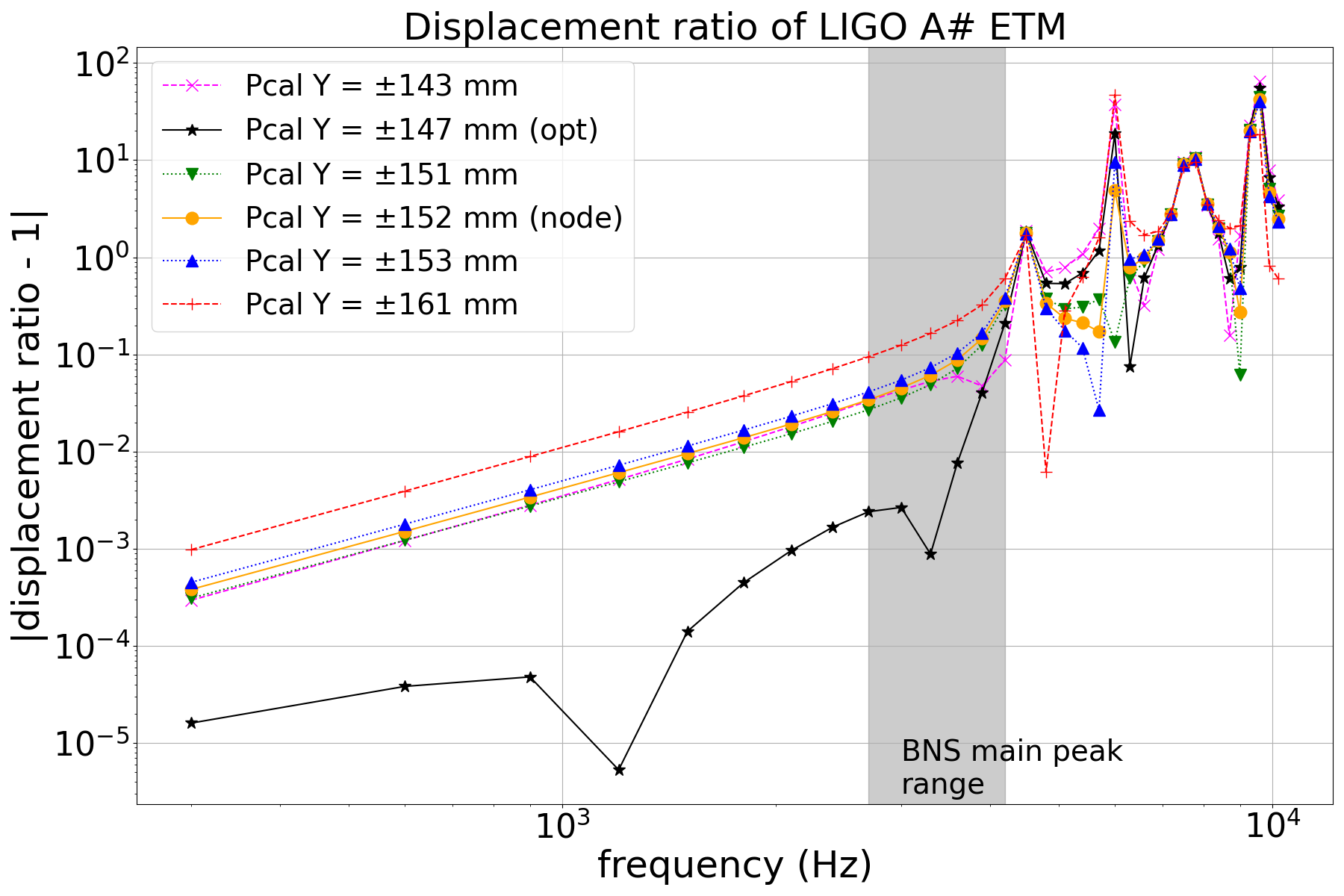}
\end{tabular}
\end{center}
\caption[]
{ \label{fig:asharp_loglog}
Discrepancy from the rigid mass motion of LIGO A\# simulated by COMSOL.
({\it Dashed Magenta Cross}) Pcal beams are 9 mm inner from the drumhead nodal radius,
({\it Solid Black Star}) Optimal position,
({\it Dotted Green Inverted Triangle}) 1 mm inner from the drumhead nodal radius,
({\it Solid Orange Circle}) Drumhead nodal radius, 
({\it Dotted Blue Triangle}) 1 mm outer from the drumhead nodal radius,
({\it Dashed Red Plus}) 9 mm outer from the drumhead nodal radius.
The frequency range from 2.7 kHz to 4.2 kHz is shaded as the region of the main ringdown peak of binary neutron star merger.
}
\end{figure}

We applied the above simulations to the configuration of KAGRA, AdVirgo, and LIGO A\#. A single-centered beam case and a two-beam case were studied for Virgo. The $\Delta d$ at 2.7 kHz and 4.2 kHz were,
\begin{itemize}
\item{KAGRA:} -2.2e-4 at 2.7 kHz and -8.1e-4 at 4.2 kHz. Optimal Pcal beam position was $\pm 75$ cm (Fig.~\ref{fig:kagra_loglog_shaded}).
\item{Virgo (one beam):} -1.9e-0 at 2.7 kHz and -4.8e-0 at 4.2 kHz. Optimal Pcal beam position was $0$ cm (Fig.~\ref{fig:virgo_center_loglog}).
\item{Virgo (two beams):} 9.7e-4 at 2.7 kHz and -4.6e-3 at 4.2 kHz. Optimal Pcal beam position was $\pm 116$ cm (Fig.~\ref{fig:virgo_twobeam_loglog}).
\item{LIGO A\#:} -2.4e-3 at 2.7 kHz and 2.1e-1 at 4.2 kHz. Optimal Pcal beam position was $\pm 147$ cm (Fig.~\ref{fig:asharp_loglog}).
\end{itemize}


\subsection{Harmonic response with measured beam offset}
\label{sect:offset}
\begin{figure}[tbp]
\begin{center}
\begin{tabular}{c}
\includegraphics[height=8.0cm]{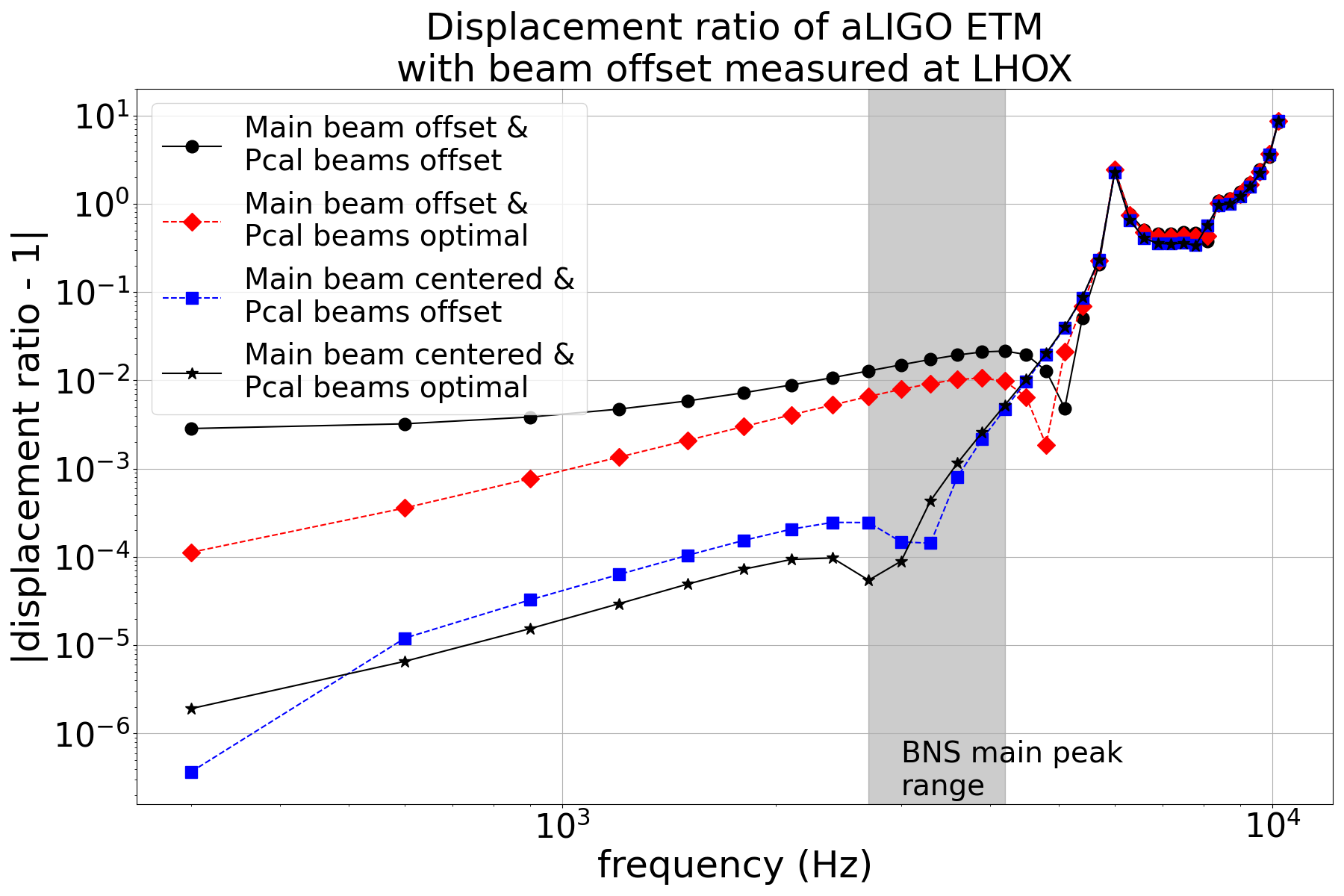}
\end{tabular}
\end{center}
\caption[]
{ \label{fig:LHO}
Discrepancy from the rigid mass motion of aLIGO's ETM.
({\it Black Solid Circle}) The main beam is at 16.2~mm rightward and 14.3~mm downward from the center, and the both Pcal beams are at 2~mm upward from their optimal positions. 
({\it Red Dashed Diamond}) The main beam is at 16.2~mm rightward and 14.3~mm downward from the center, and the both Pcal beams are at their optimal positions.
({\it Blue Dashed Square}) The main beam is at the center of the mirror, and the both Pcal beams are at 2~mm upward from their optimal positions.
({\it Black Solid Star}) The main beam is at the center of the mirror, and the Pcal beams are at their optimal positions.
The frequency range from 2.7~kHz to 4.2~kHz is shaded as the region of the main ringdown peak of binary neutron star merger.
}
\end{figure}

Finally we simulated the cases that the Pcal beams and the main interferometer beam have offset from their optimal positions. Offset of the main beam was the same as Sec.~\ref{sect:cylsus}, which was 16.2~mm to the right and 14.3~mm below the center, and the offset of the Pcal beams is estimated to be less than 2 mm~\cite{LHO_mainbeam}. In order to evaluate the worst case, here we considered a case where the both Pcal beams are vertically moved for 2 mm in the opposite way of the main beam offset because the butterfly mode is larger when the Pcal beams have vertical offset than horizontal due to asymmetric shape of the ETM. We compared it with results of the case that only the Pcal beams or only the main beam have offset. The $\Delta d$ with all of the Pcal beams and the main beam having offsets was -1.3e-2 at 2.7 kHz and -2.1e-2 at 4.2 kHz as shown in Fig.~\ref{fig:LHO}. This implies that the realistic beam offsets in LIGO would result in a systematic error of over 1\% at 2.1 kHz and above, which is the important range in the studies of binary neutron stars in future generations of GW observations. This magnitude is comparable with the systematic error estimated for the Pcal-induced ETM rotation, which was 0.41\% at LHO~\cite{ligo_o4_calibration_dcc}.

\subsection{Response in wide frequency range}
\begin{figure}[tbp]
\begin{center}
\begin{tabular}{c}
\includegraphics[height=8.0cm]{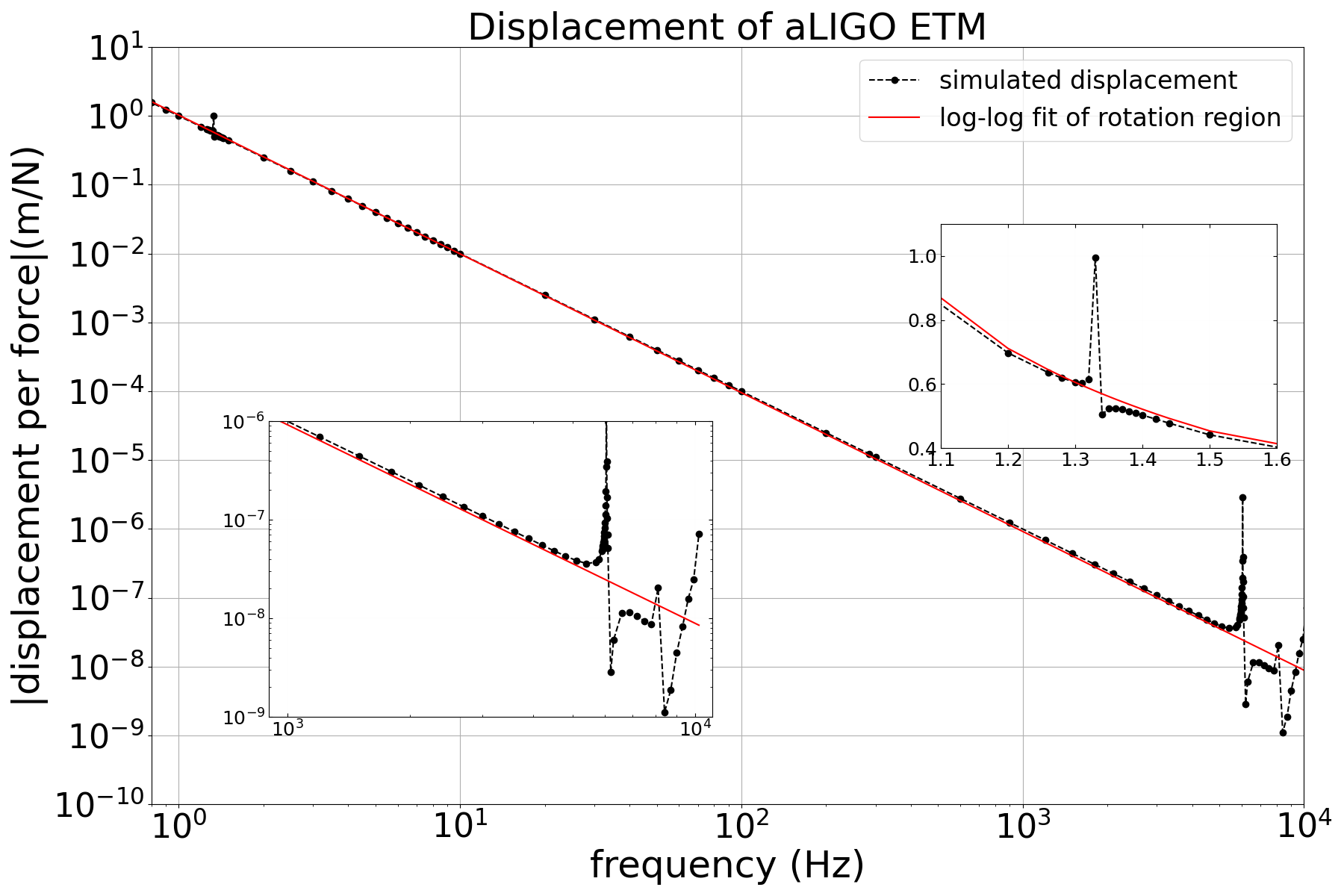}
\end{tabular}
\end{center}
\caption[]
{ \label{fig:bulkrotbudget}
Surface displacement of aLIGO's ETM normalized by the Pcal laser power. Black dashed line is the simulated displacement while the red solid line is a log-log fit of the 1-10~Hz data. 
({\it Upper right subplot}) the resonant peak of the pitch rotation mode around 1.3~Hz.
({\it Lower left subplot}) the bulk deformation peaks at kHz region.
}
\end{figure}

Figure~\ref{fig:bulkrotbudget} shows displacement $x$ simulated using aLIGO's two-staged suspension model with beam offsets of the main beam and the Pcal beams. The position of the main beam reproduced the measurement in LHO, and both Pcal beams were at 2~mm upward from their optimal positions. The $1/M\omega^2$ structure dominated above 1~Hz where the resonant peaks of pitch rotational modes exist. The bulk deformation rises as a discrepancy from the $1/M\omega^2$ structure above 1500 Hz. Modal analysis inferred that the violin mode of the fibers is excited altogether with the bulk modes of the mirror.

\begin{figure}[tbp]
\begin{center}
\begin{tabular}{c}
\includegraphics[height=8.0cm]{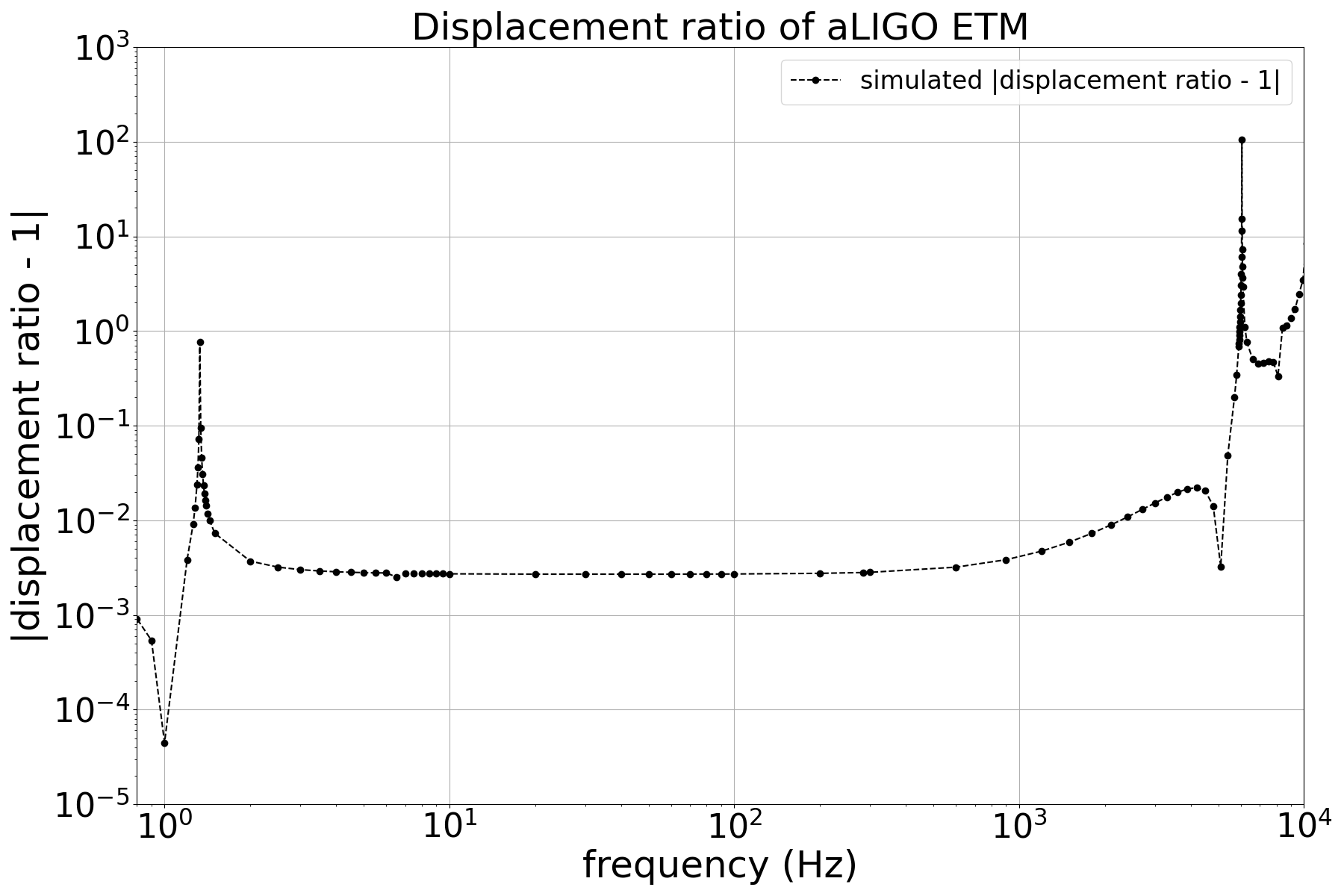}
\end{tabular}
\end{center}
\caption[]
{ \label{fig:bulkrotbudget_disp}
Displacement ratio of aLIGO's ETM in the wide frequency range. 
}
\end{figure}

Figure~\ref{fig:bulkrotbudget_disp} shows the displacement ratio calculated by normalizing the main curve in Fig.~
\ref{fig:bulkrotbudget} by $1/M\omega^2$. It shows the discrepancy from the rigid-mass motion which is important from the perspective of calibration. The displacement ratio is flat in the range from a few Hz to a few hundred Hz, but it start rising around 300~Hz due to the internal modes.

\section{Discussion}
In this study, the formulation of transfer functions was summarized in Section 2. We made an assumption that the interaction between different modes is small. Based on this assumption, we introduced the concept that each fluctuation is additive. This conclusion is inferred from the independence of the Lagrangian. This formulation is applicable even when Pcal beams are multiple, suggesting its applicability in scenarios involving injection at multiple points in the future and in systems like current Virgo. The generality of this formulation expands the framework for approaching problems, making it applicable to a wider range of situations. Thus, it will provide a foundation for offering more flexible and effective methodologies in future research and practice.

In previous research, De Lillo {\em et al.} (2017) simulated bulk deformation of aLIGO ETM, and Karki (2019) conducted a comparison between experimental data and models~\cite{Nicola_pcal}\cite{sudarshan_phd}. Their experimental data supported the simulation results. While the actual measurements are limited by the sensitivity of the detection apparatus to a certain frequency range, the revelation of broad frequency response through simulations is significant. These results indicate that simulations appropriately reproduce realistic conditions of experiments and serve as valuable tools for deepening our understanding behind the measured response. Furthermore, the agreement with experimental data confirms the reliability of the model, suggesting that such comparisons would be beneficial in future experiments and simulations.

In this study, we compared responses of LIGO ETM simulated by ANSYS and COMSOL. Consistency between these two software within 1\% validates our simulation methods and results. The small discrepancy could be from a difference of surface integration method in two analyses. 

The two-staged suspension model enables us to simulate more realistic system and include rotational effect around fiber-attachment points and suspension supports as well as the bulk deformation effect. As the model becomes complicated, we see more eigenfrequencies of the violin modes of the suspension fibers. According to the comparison of displacement ratio of free-mass and suspension models, their impact to calibration is expected to be less than 1\% if we choose frequencies sufficiently apart from the resonance.

We neglected phase of the displacement in this study. Although the phase difference between translational, rotational, and internal modes are summed in the surface displacement, the phase delay to the external force is not contained. From the perspective of calibration, it has no significant effect at frequencies sufficiently far from the resonance. It becomes non-negligible when we consider a response near the resonant frequencies.

Study of KAGRA's ETM showed advantage of material in a perspective of moving the resonant frequency high away from the observation range. On the other hand, it was revealed that the optimal Pcal beam points are away from the nodal point of the drumhead mode. This is due to the asymmetry in the positioning of the ears used to fix the mirrors. Since the supporting structure used to fix the mirrors is not evenly distributed vertically, asymmetry occurs in the behavior of the mode shapes. The asymmetry of the butterfly mode results in the residual butterfly mode even when the main beam is aligned at the center. Therefore, the optimal point, where the contributions of the butterfly mode and the drumhead mode are minimized, has to be identified.

In Virgo, Pcal is currently injected into the center in the mirror. However, center injection can excite the drumhead mode, requiring sophisticated modeling of the mirror response. However, the transfer function becomes to a simpler form of $1/f^2$ when Pcal is injected at two points. In this study, we evaluated the effect of bulk deformation in these two cases. We compared the cases with Pcal beams positioned at the center and at vertically-symmetric two points. Center placement raises concerns of exciting the drumhead mode, necessitating complex modeling to address it. On the other hand, when the Pcal beams are placed at two points, the bulk deformation effect was significantly reduced due to suppression of the drumhead mode.

We also conducted analyses regarding the LIGO A\# project. In LIGO A\#, it is planned to increase the size of ETM to 100 kg to suppress statistical errors. We examined the expected systematic errors due to this change. The estimated eigenfrequency of butterfly mode was 4282.1 Hz, which is close to the frequency of the characteristic peak in the ringdown phase of binary neutron star mergers. Additionally, due to the influence of this peak, it becomes apparent that a $1/f^2$ transfer function cannot be used across a wide frequency range. The frequency range where this transfer function can be applied with 1\% accuracy is below 1500 Hz, which is not enough to cover the science range of binary neutron stars and burst events. These suggest a need for reducing systematic errors. To address these issues, it is necessary to suppress the butterfly mode caused by Pcal injection. Injecting three or more Pcal beams could realize it. Independently controlled multiple injections can also be applied to a photon-pressure actuator to control the mirror without coupling to environmental magnetic field.

\section{Summary}
\label{sect:summary}
We performed modal analysis and harmonic response analysis with ETMs of aLIGO, AdVirgo, KAGRA, and LIGO A\#. We checked consistency within 1\% between ANSYS and COMSOL to validate simulations. We also confirmed that a free-mass ETM model can provide consistent results with more realistic suspension models within 1\%. Then we evaluated displacement ratio to the rigid-mass motion with the various cases of misalignment of Pcal beams and the main IFO beam. Measured main beam offset in LHO and conservatively estimated Pcal beams offset can induce percent-order discrepancy from the $1/f^2$ transfer function through the bulk deformation in the kHz region.

\acknowledgments     
This research has made use of data or software obtained from the Gravitational Wave Open Science Center (gwosc.org), a service of the LIGO Scientific Collaboration, the Virgo Collaboration, and KAGRA. This material is based upon work supported by NSF's LIGO Laboratory which is a major facility fully funded by the National Science Foundation, as well as the Science and Technology Facilities Council (STFC) of the United Kingdom, the Max-Planck-Society (MPS), and the State of Niedersachsen/Germany for support of the construction of Advanced LIGO and construction and operation of the GEO600 detector. Additional support for Advanced LIGO was provided by the Australian Research Council. Virgo is funded, through the European Gravitational Observatory (EGO), by the French Centre National de Recherche Scientifique (CNRS), the Italian Istituto Nazionale di Fisica Nucleare (INFN) and the Dutch Nikhef, with contributions by institutions from Belgium, Germany, Greece, Hungary, Ireland, Japan, Monaco, Poland, Portugal, Spain. KAGRA is supported by Ministry of Education, Culture, Sports, Science and Technology (MEXT), Japan Society for the Promotion of Science (JSPS) in Japan; National Research Foundation (NRF) and Ministry of Science and ICT (MSIT) in Korea; Academia Sinica (AS) and National Science and Technology Council (NSTC) in Taiwan.
\bibliography{report}   
\bibliographystyle{chronosbib} 

\end{document}